\documentclass[journal]{IEEEtran}

\usepackage{cite}

\usepackage{graphicx}

\usepackage{amsmath}

\usepackage{array}

\usepackage{url}

\begin{document}

\title{A Compact XOR Gate Implemented With a Single Straintronic Magnetic Tunnel Junction}
%
%
%

\author{Supriyo~Bandyopadhyay,~\IEEEmembership{Life~Fellow,~IEEE} \\
\bigskip

\thanks{Manuscript received ...; revised ... 
 ({\it Corresponding author: Supriyo Bandyopadhyay})}

\thanks{The author is with the Department
of Electrical and Computer Engineering, Virginia Commonwealth University, Richmond, VA, 23284 USA. (email:  sbandy@vcu.edu)}} 

\maketitle

\begin{abstract}
 The XOR Boolean logic gate is widely used in many applications such as encryption (XOR ciphers), binary addition (half- and full-adders), error detection (parity bits), etc. but is challenging to construct because of its demanding conditional dynamics. It typically  requires multiple logic switches or other types of gates, which results in a large gate footprint and low logic density. Here, we present the design of an XOR gate with a single straintronic magnetic tunnel junction which reduces the footprint dramatically. Such a gate is non-volatile and hence suitable for non-von-Neumann architectures, processor-in-memory, etc. The switching time of the gate is $\sim$200 ps and the energy dissipation per gate operation is $\sim$225 aJ. Cascading of successive stages is accomplished via a CMOS device which plays no role in the gate dynamics but is needed for gain to provide logic level restoration, fan-out and isolation between input and output. This 1 MTJ - 1 CMOS design has an energy dissipation that is an order of magnitude smaller than what has been reported for traditional all-transistor XOR designs.
\end{abstract}


%
\IEEEpeerreviewmaketitle

\section{Introduction}

\IEEEPARstart{A}{n} XOR gate is a Boolean gate that produces a low output state (output bit = 0) when the two input bits are identical and a high output state (output bit = 1) when the two input bits are logic complements of each other. This dynamics is challenging  to implement and  therefore the XOR usually requires more primitive devices (logic switches) to construct than most other gates (e.g., 4-6 NANDs are required to implement an XOR), which results in a large footprint. Here, we remedy this problem by using a single magnetic tunnel junction (MTJ) to realize an XOR gate.  The MTJ is switched with {\it electrically generated mechanical strain} which naturally enables the XOR functionality. This would not have been possible had the MTJ been switched with conventional spin transfer torque \cite{ralph}, or spin-orbit torque \cite{shao} or voltage controlled magnetic anisotropy \cite{khalili}. The designed XOR gate has the added advantage of being non-volatile because of the use of a magnetic element as the logic processor.

\section{Gate Design}

\begin{figure}[hbt!]
\centering
\includegraphics[width=0.46\textwidth]{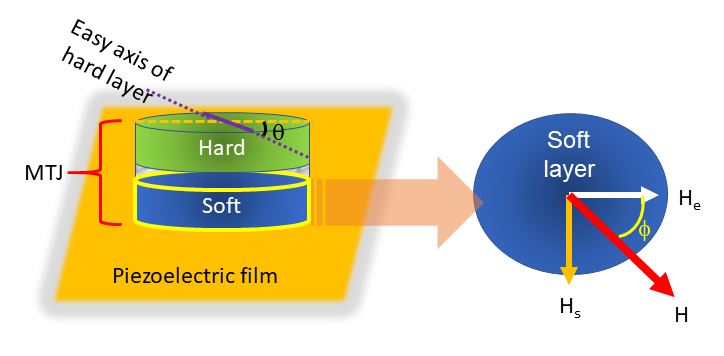}
\caption{The net effective magnetic field in the elliptical magnetostrictive soft layer of a magnetic tunnel junction subjected to an external magnetic field and uniaxial strain along the major axis. The angle subtended by the net magnetic field with the major axis of the soft layer is $\phi$. The easy axis of the hard layer subtends an angle $\theta$ with its own major axis.}
\label{fig:net field}
\end{figure}

To understand the basic concept behind the XOR design, consider an MTJ whose elliptical soft layer is magnetostrictive and placed in elastic contact with an underlying piezoelectric layer [see Fig. \ref{fig:net field}]. If a voltage is applied across the piezoelectric using a suitable electrode configuration \cite{cui}, then biaxial strain is generated in it, which is transferred to the soft layer of the MTJ and rotates its magnetization from the major (easy) axis towards the minor (hard) axis owing to the inverse magnetostriction (or Villari) effect. This will happen only if the product of the magnetostriction and the strain component along the major axis has a negative sign (or equivalently the product of the magnetostriction and the strain component along the minor axis has a positive sign) \cite{kuntal1,kuntal2,APR,monograph}. Such a rotation  will change the MTJ resistance. Very little energy is dissipated in this process \cite{kuntal1,kuntal2,APR,monograph}. Strain mediated switching of MTJs have been demonstrated, albeit with thick piezoelectric layers that compromised the energy efficiency \cite{li,zhao}.

Here, the strain creates an effective magnetic field {\bf H$_s$} along the minor axis, which is why the magnetization rotates towards it. Assume now that an external magnetic field {\bf H$_e$} is applied along the major axis of the ellipse. When both {\bf H$_s$} and {\bf H$_e$} are present, the magnetization of the soft layer will settle along the net magnetic field given by ${\bf H} = {\bf H_e} + {\bf H_s}$ as shown in Fig. \ref{fig:net field}.

If we keep on increasing the stress on the soft layer, the magnitude of {\bf H$_s$} will increase and the soft layer's magnetization will increasingly move towards the minor axis. Let us say that at any given stress magnitude, the acute angle subtended by {\bf H} with the soft layer’s major axis is $\phi$ as shown, where $\phi = tan^{-1}\left ({\bf H}_s/{\bf H}_e \right )$. We also assume that the hard layer’s easy axis subtends an angle $\theta$ with its own major axis as shown in Fig. \ref{fig:net field}. The resistance of the MTJ is given by $R_{MTJ} = R_P + \left ( R_{AP} - R_P \right )/2 \left [ 1 - cos(\theta - \phi) \right ]$, where $R_P$ is the lowest resistance corresponding to the magnetizations of the hard and soft layers being parallel and $R_{AP}$ is the highest resistance corresponding to them being antiparallel. Clearly, the MTJ resistance will be lowest (i.e.,  $R_P$) when $\theta = \phi$.   If we under-stress the soft layer, then $\theta > \phi$ and the MTJ resistance will be higher. If we over-stress, then $\theta < \phi$ and again the MTJ resistance will be higher. This feature is leveraged to implement the XOR functionality.

\begin{figure*}[t!]
\centering
\includegraphics[width=0.91\textwidth]{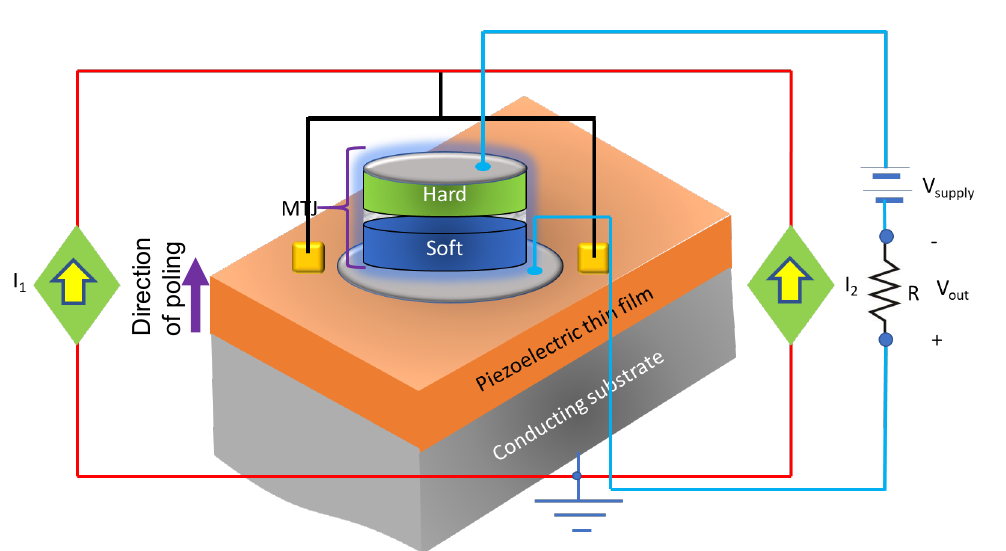}
\caption{Design of an XOR gate with a single straintronic MTJ. The two inputs are encoded in the currents $I_1$ and $I_2$, whereas the output is encoded in the voltage $V_{out}$.}
\label{fig:XOR}
\end{figure*}

Consider now the construct in Fig. \ref{fig:XOR}. It consists of a MTJ with a magnetostrictive soft layer in elastic contact with a poled piezoelectric thin film deposited on a conducting substrate. Two electrodes are delineated on the piezoelectric's surface such that the spacing between the edge of the MTJ's soft layer and that of the nearest electrode is of the same order as that of the piezoelectric film thickness \cite{cui}. The two electrodes are shorted together and the back of the conducting substrate is grounded. This is the design of the XOR gate where the two input bits are encoded in the currents $I_1$ and $I_2$, and the output is encoded in the voltage $V_{out}$ dropped across the resistor R placed in series with an external voltage source $V_{supply}$. 

\begin{figure}[hbt!]
\centering
\includegraphics[width=0.46\textwidth]{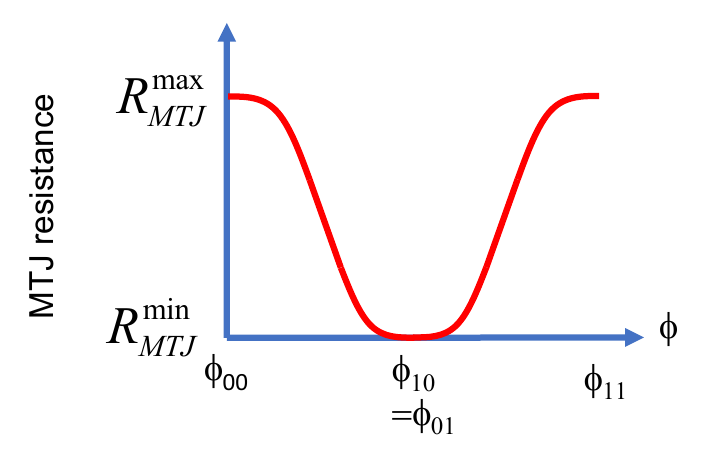}
\caption{Schematic representation of the resistance of the MTJ as a function of the strain-induced rotation angle $\phi$.}
\label{fig:resistance}
\end{figure}

The input currents $I_1$ and $I_2$ produce a voltage drop of $V_p$ across the piezoelectric layer, where $V_p \approx R_{piezo} \left [ I_1 + I_2 \right]$ and $R_{piezo}$ is the resistance of the piezoelectric layer underneath the contact pad. The voltage $V_p$ generates a vertical electric field ${\cal E}$ across the piezoelectric, which results in  a biaxial strain in the piezoelectric layer underneath the MTJ \cite{cui}, which is transferred to the soft layer and generates the magnetic field {\bf $H_s$} along the minor axis if the polarity of $V_p$ is chosen to fulfill the condition that the product of the magnetostriction and strain component along the soft layer's major axis has a negative sign (or equivalently the product of the magnetostriction and the strain component along the minor axis has a positive sign). The magnetic field {\bf $H_s$} rotates the soft layer's magnetization through an angle $\phi$ away from the major axis. The strain component along the major axis within the piezoelectric is approximately $d_{33}{\cal E}$, where $d_{33}$ is the piezoelectric coefficient of the piezoelectric material. The magnitude of {\bf $H_s$} is proportional to ${\cal E}$ and hence $V_p$. Thus, increasing $V_p$ increases $\phi$. 

\paragraph{Gate dynamics}: When both input currents $I_1$ and $I_2$ are zero, the total voltage dropped across the piezoelectric layer is $V_p$ = 0. Hence no strain is generated and $\phi = \phi_{00}$ = 0. When either voltage is positive, but not both, the voltage drop is $V_p \approx R_{piezo} I_{in}$, where $I_{in}$ is the magnitude of $I_1$ or $I_2$. This generates a strain that rotates the magnetization of the soft layer through an angle $\phi_{10} = \phi_{01}$ away from the major axis. We will make $\theta = \phi_{10} = \phi_{01}$ in order to realize the XOR functionality. Finally, when  both $I_1$ and $I_2$ are positive, $V_p \approx 2  R_{piezo} I_{in}$. This generates more strain and makes the soft layer's magnetization rotate through an angle $\phi_{11} > \phi_{01}, \phi_{10}, \theta$. The resistance of the MTJ as a function of the angle $\phi$ is shown schematically in Fig.\ref{fig:resistance}. Ideally, we will want $\theta - \phi_{00} = \phi_{11} - \theta$. Since $\phi_{00}$ = 0, we get that $\theta = \phi_{11}/2$. Thus, the angle $\theta$ has to be chosen to be one-half of the maximum rotation of the soft layer's magnetization that takes place when both inputs are 1.

\paragraph{XOR functionality}: The voltage $v_{out}$ is the gate's output. It is given by $v_{out} = V_{supply} R/\left (R + R_{MTJ} \right ) \approx V_{supply} R/R_{MTJ}$ ($R_{MTJ} \gg R$), where $R_{MTJ}$ is the MTJ resistance. Hence, $v_{out}$ is small when $R_{MTJ}$ is large and large when $R_{MTJ}$ is small. Since $R_{MTJ}$  is small only when one input voltage is positive and the other is zero, we obtain the truth table in Table I, which is that of an XOR gate. Hence, this construction acts as an XOR gate.

\hspace{1cm}
\begin{table}
    \caption{Truth table of the logic gate shown}
    \centering
    \begin{tabular}{|c|c|c|}
    \hline
    Input 1 ($I_1$) & Input 2 ($I_2$) & Output ($v_{out})$ \\
    \hline
    && \\
    0 & 0 & 0 \\
    1 & 0 & 1 \\
    0 & 1 & 1 \\
    1 & 1 & 0 \\
    \hline
    \end{tabular}
\end{table}
\hspace{1cm}

\section{Appendix}

\subsection{Standby power dissipation}
The supply voltage $V_{supply}$ is an independent voltage source and determines $v_{out}$. It must be large enough to make $v_{out}$ larger than the input voltage pulse amplitudes that produce $I_1$ and $I_2$. This source also causes standby power dissipation but that can be made arbitrailty small by making the MTJ resistance very large even when the two inputs are logic complements of each other. There are many ways of accomplishing that, such as by choosing a small cross sectional area of the MTJ or a thick spacer layer. 

\begin{figure*}[hbt!]
\centering
\includegraphics[width=0.91\textwidth]{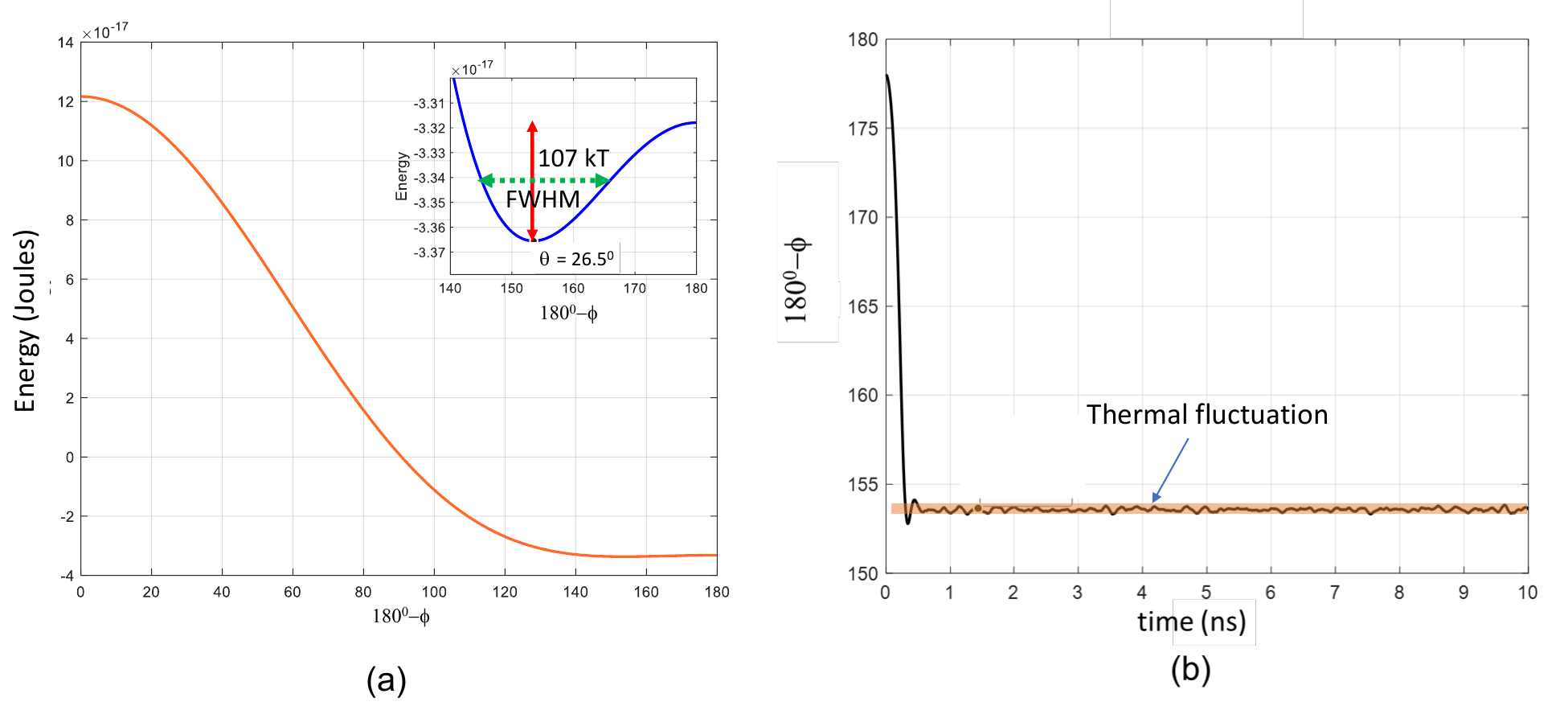}
\caption{(a) The potential well surrounding the critical angle. (b) The fluctuation in the critical angle due to thermal noise at room temperature is a mere $\sim$ 0.2 degrees. Reproduced from \cite{rahnuma} with permission of the IEEE.}
\label{fig:data}
\end{figure*}

\subsection{Stability of $\phi$}
The angle $\phi$, for any combination of the two input currents, must be {\it stable} and not drift or fluctuate through large angles owing to thermal noise. This is required for gate reliability. The magnetization dynamics under strain in a similar system was investigated in ref. [\citenum{rahnuma}] using Landau-Lifshitz-Gilbert equations for an MTJ with characteristics given in Table II and it was found that at any value of $\phi$, the potential well that surrounds it to make it stable is both wide and deep. This is shown in Fig. \ref{fig:data}. The depth of the well is 107 kT at room temperature and the full width at half maximum (FWHM) is 20 degrees which is 22\% of the phase space of 90 degrees for the angle $\phi$. The former ensures that thermal fluctuations cannot disturb $\phi$ and the latter ensures that there is little possibility of the angle overshooting or undershooting the critical angle since the potential well is broad. These characteristics are very encouraging since the basin surrounding the critical angle is a deep and wide attractor.

\hspace{1cm}
\begin{table}
    \caption{Parameters for the soft layer of the MTJ simulated in Fig. \ref{fig:data}}
    \centering
    \begin{tabular}{|c|c|}
    Major axis dimension & 800 nm\\
    Minor axis dimension & 700 nm \\
    Thickness & 2.2 nm \\
    Saturation magnetization & 8.5 $\times$ 10$^5$ A/m\\
    Magnetic field $H_e$ & 1000 Oe \\
    Gilbert damping constant & 0.1\\
    Saturation magnetostriction & 600 ppm \\
    Young's modulus & 120 GPa \\
    Piezoelectric coefficient d$_{33}$ & 1.5$\times$10$^{-9}$ C/m \\
    Piezoelectric layer thickness & 1 $\mu$m \\
    \end{tabular}
\end{table}
\hspace{1cm}

\subsection{Switching speed} 
Fig. \ref{fig:data}(b) shows that once the stress is turned on, the stable state is reached in $\sim$200 ps which portends relatively fast switching. Ferromagnets, unfortunately, do not usually switch very fast while antiferromagnets switch much faster but do not have dipole moments (non-collinear ones may have octuple moments) and hence are not preferred for logic applications since the magnetic signal strength will be very weak. It may be possible to use time varying strain in conjunction with pulsed current generating spin-orbit or spin-transfer torque to increase switching speed of the ferromagnetic MTJ considered here and reduce switching error probability \cite{ayan1} at the same time, but that study is outside the scope of this work. Experimental studies of dual mode switching of MTJs (time varying strain + spin transfer torque) have been carried out recently \cite{zink}, albeit not under optimum conditions.

There may be scenarios where fast switching is the primary consideration. In those cases, one can use ferrimagnets near the compensation point instead of a ferromagnetic soft layer. That can increase the switching speed at the cost of a lower magnetic moment and hence lower signal strength.

\subsection{Energy dissipation}
The major component of the energy dissipated during switching is $CV^2$ where $C$ is the capacitance of the gate pads used to apply voltage on the piezoelectric to generate strain and $V$ is the voltage needed to generate the strain. The voltage needed to generate a given amount of strain depends on the thickness of the piezoelectric layer. Lead-zirconate-titanate (PZT) films grown on 10 $\mu$m thick mica substrates have exhibited a piezoelectric $d_{33}$ coefficient of 160 pC/N when the PZT layer thickness is 50 nm \cite{zhang}. The strain generated $\epsilon$ in it by an electric field ${\cal E}$ is 
\begin{equation}
    \epsilon  \approx d_{33} {\cal E}.
    \label{strain}
\end{equation}

\begin{figure*}[hbt!]
\centering
\includegraphics[width=0.91\textwidth]{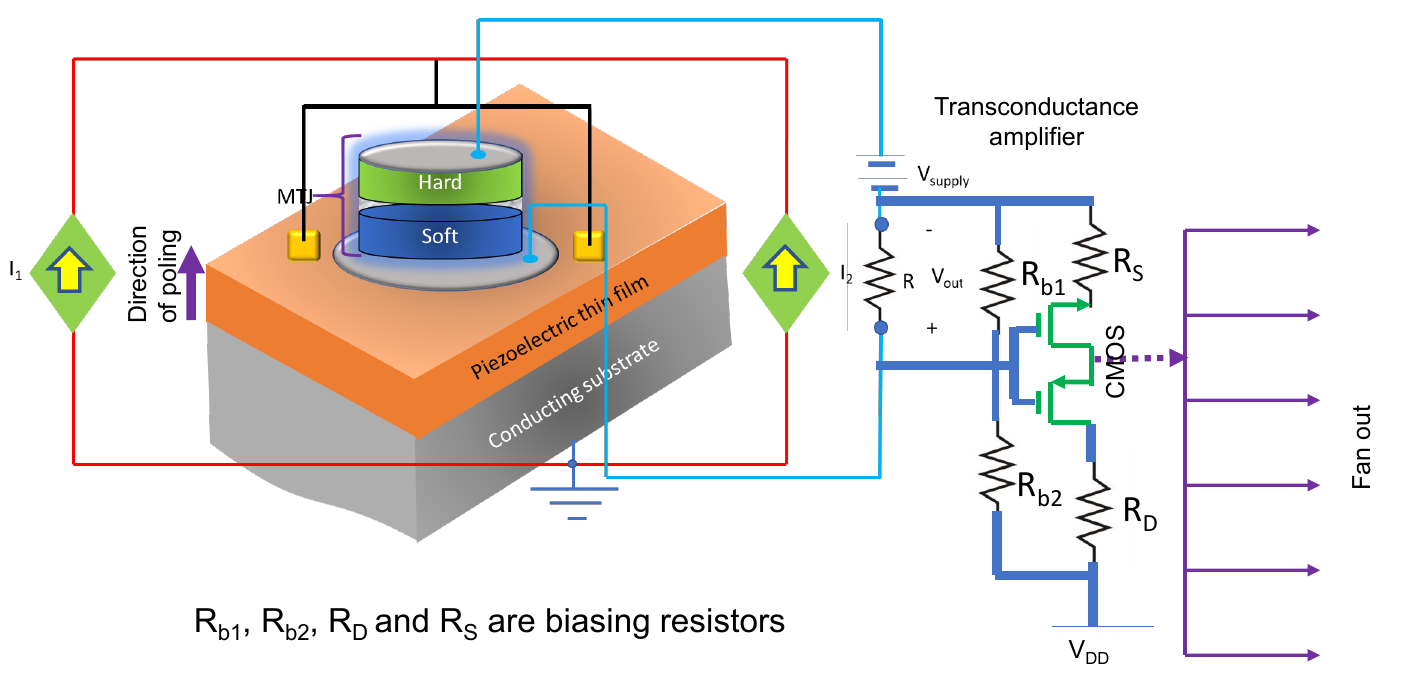}
\caption{The concatenation scheme for providing gain for logic level restoration and fan-out.}
\label{fig:concatenation}
\end{figure*}

We will choose the soft layer materials to be Terfenol-D because of its large magnetostriction. Ref. [\citenum{kuntal2}] showed that 30 MPa of stress is more than enough to rotate the magnetization of a Terfenol-D soft layer (the material considered in ref. [\citenum{rahnuma}]) through 90$^{\circ}$ at room temperature in 200 ps or less. Here, $\phi_{11} < 90^{\circ}$ and hence we will need less than 30 MPa of stress. The Young's modulus of Terfenol-D is about 60 GPa. Hence, the maximum strain that will need to be generated is 30 MPa/60 GPa = 5$\times$10$^{-4}$. From Equation (\ref{strain}), we get that the electric field needed to generate this much strain in the  piezoelectric thin film is about 3 MV/m. Therefore, the voltage $V$ needed to generate this electric field across a 50 nm thin film of PZT is 3 MV/m $\times$ 50 nm = $\sim$150 mV. 

Next, we estimate the capacitance of the gate pads (electrodes). Sub-micron gate pad dimensions are now common \cite{contacts}. We will assume a pad area of 1 $\mu$m $\times$ 1 $\mu$m. The relative dielectric constant of PZT thin films of around 50 nm thickness is reported to range from 30-130 \cite{boni} and we will choose a value of 60. Hence the gate capacitance is about 
$C = 60 \times 8.854 \times 10^{-12} \times 10^{-12}/(50 \times 10^{-9})$ = 10 fF. Therefore, the energy dissipation per gate operation is $CV^2$ $\approx$ 225 aJ, which is much better than that of optimized conventional designs \cite{saravanan}. We also note that the room temperature thermal noise voltage appearing at the gate pad is $\sqrt{kT/C}$ = 646 $\mu$V, which is much less than 150 mV. Hence, the gate has very strong noise immunity and logic bits are unlikely to be corrupted by noise.

We point out that there are other declamped piezoelectric thin films that have exhibited high piezoelectric coefficients \cite{ryan}, such as patterned rhombohedral \{001\}-oriented 70/30 lead magnesium niobate - lead titanate (PMN-PT) thin films of 300 nm thickness grown on platinized Si substrates. They have exhibited a $d_{33}$ of $\sim$160 pC/N and may be easier to synthesize, but their relative permittivity is higher, which will result in somewhat higher gate capacitance and hence energy dissipation. 

\subsection{Concatenation, cascading, fan-out and logic restoration}

For the sake of logic restoration and fan-out, successive stages have to be cascaded through a CMOS which provides isolation between input and output as well as gain for logic level restoration (including current gain for fan-out). The concatenation scheme is shown in Fig. \ref{fig:concatenation}. The CMOS acts as a transconductance amplifier. It adds an energy cost, but modern CMOS can dissipate as low as 30 aJ to switch \cite{suman} and hence the energy cost is tolerable. 

\section{Conclusion}

An XOR gate typically takes 6-12 transistors to build in conventional designs \cite{wiki}. Here, we have presented a design that utilizes a single magnetic tunnel junction, resulting in much reduced footprint. The switching speed is $\sim$200 ps and the energy dissipation per gate operation is $\sim$225 aJ, plus an addition $\sim$35 aJ for the CNOS, making the total energy dissipation about 260 aJ. This is at least an order of magnitude smaller than what is reported for conventional all-CMOS designs. The other significant advantage is that the gate is {\it non-volatile} since it is synthesized from a magnetic tunnel junction (the volatility of the CMOS is immaterial since it is only used for cascading). That makes it amenable to non-von-Neumann architectures, processor-in-memory circuits, etc. and is also better suited for edge computing and internet of things.

\bigskip

\noindent {\bf Acknowledgment}
The author is indebted to Prof. Avik Ghosh of the University of Virginia for helpful discussions.


\begin{thebibliography}{1}



\bibitem{ralph}
D. C. Ralph and M. D. Stiles, ``Spin transfer torques'', {\it J. Magn. Magn. Mater.}, vol. 320, 1190-1216 (2008). https://doi.org/10.1016/j.jmmm.2007.12.019.
\bibitem{shao}
Q. Shao, et al., ``Roadmap of spin-orbit torques'', {\it IEEE Trans. Magn.}, vol. 57, Art. 800439 (2021). https://doi.org/10.1109/TMAG.2021.3078583
\bibitem{khalili}
Y. Shao and P. K. Amiri, ``Progress and application perspectives of voltage-controlled magnetic tunnel junctions'', {\it Adv. Mater. Technol.}, vol. 8, Art. 2300676 (2023).  https://doi.org/10.1002/admt.202300676.
\bibitem{cui}
J. Cui, J. L. Hockel, P. K. Nordeen, D. M. Pisani, C.-Y Liang, G. P. Carman and C. S. Lynch, 
``A method to control magnetism in individual strain-mediated magnetoelectric islands'', {\it Appl. Phys. Lett.}, vol. 103, Art. 232905 (2013). https://doi.org/10.1063/1.4838216.
\bibitem{kuntal1}
K. Roy, S. Bandyopadhyay and J. Atulasimha, ``Hybrid spintronics and straintronics: A magnetic technology for ultra low energy computing and signal processing'', {\it Appl. Phys. Lett.}, vol. 99, Art. 063108 (2011). https://doi.org/10.1063/1.3624900.
\bibitem{kuntal2}
K. Roy, S. Bandyopadhyay and J. Atulasimha, ``Energy dissipation and switching delay in stress-induced switching of multiferroic nanomagnets in the presence of thermal fluctuations'', {\it J. Appl. Phys.}, vol. 112, Art. 023914 (2012). https://doi.org/10.1063/1.4737792
\bibitem{APR}
S. Bandyopadhyay, J. Atulasimha and A. Barman, ``Magnetic straintronics: Manipulating the magnetization of magnetostrictive nanomagnets with strain for energy-efficient applications'', {\it Appl. Phys. Rev.}, vol. 8, Art. 041323 (2021). https://doi.org/10.1063/5.0062993.
\bibitem{monograph}
S. Bandyopadhyay, {\it Magnetic Straintronics:
An Energy-Efficient Hardware Paradigm for Digital and Analog Information Processing} (Springer-Nature, Cham, Switzerland, 2022).
\bibitem{li}
P. Li, A. Chen, D. Li, Y. Zhao, S. Zhang, L. Yang, Y. Liu, M. Zhu, H. Zhang,
and X. Han, ``Electric field manipulation of magnetization rotation and tunneling magnetoresistance of magnetic tunnel junctions at room temperature'', {\it Adv. Mater.}, vol. 26, 4320-4325 (2014). https://doi.org/10.1002/adma.201400617.
\bibitem{zhao}
Z. Zhao, M. Jamali, N. D’Souza, D. Zhang, S. Bandyopadhyay, J. Atulasimha,
and J.-P. Wang, ``Giant voltage manipulation of MgO-based magnetic tunnel
junctions via localized anisotropic strain: A potential
pathway to ultra-energy-efficient memory technology'', {\it Appl. Phys. Lett.}, vol. 109, Art. 092403 (2016). https://doi.org/10.1063/1.4961670.
\bibitem{rahnuma}
R. Rahman and S. Bandyopadhyay, ``A nonvolatile all-spin nonbinary matrix multiplier: an efficient hardware accelerator for machine learning'', vol. 69, 7120-7127 (2022). 
https://doi.org/10.1109/TED.2022.3214167.
\bibitem{ayan1}
A. K. Biswas, S. Bandyopadhyay and J. Atulasimha, ``Acoustically assisted spin-transfer-torque switching of nanomagnets: An energy-efficient hybrid writing scheme for non-volatile memory'', {\it Appl. Phys. Lett.}, vol. 103, Art. 232401 (2013). http://dx.doi.org/10.1063/1.4838661.
\bibitem{zink}
B. Zink, B. Ma, D. Zhang, D. Bhattacharya, M. A. Abeed, S. Bandyopadhyay, J. Atulasimha and J-P Wang, ``Influence of surface acoustic wave (SAW) on nanoscale in-plane magnetic tunnel junctions'', {\it AIP Advances}, vol. 14, Art. 025104 (2024). https://doi.org/10.1063/9.0000823.
\bibitem{zhang}
R. Zhang, Y. Ding, N. Liu, W. Tang, Y. Wang, Y. Yang, Y. Wang and G. Yuan, ``High macroscopic piezoelectric $d_{33}$ of the nm-thick flexible PZT ferroelectric film'', {\it J. Mater. Sci: Mater. Elec.}, vol. 35, Art. 298 (2024). https://doi.org/10.1007/s10854-024-12040-6.
\bibitem{contacts}
K. Daffe, J. Marzouk, C. Boyaval, G. Dambrine, K. Hadaddi and S. Arscott, ``A comparison of pad metallization in miniaturized microfabricated silicon microcantilever-based wafer probes for low contact force low skate on-wafer measurements'', {\it J. Micromechanics Microengineering}, vol. 32, Art. 015007 (2022). https://doi.org/10.1088/1361-6439/ac3cd7.
\bibitem{boni}
G. A. Boni, C. F. Chirila, L. Hrib, R. Negrea, L. D. Filip, I. Pintilie and L. Pintilie, ``Low value for the static background dielectric constant in epitaxial PZT thin films'', {\it Sci. Rep.}, vol. 9, Art. 14698 (2019). https://doi.org/10.1038/s41598-019-51312-8.
\bibitem{saravanan}
Saravanan P and Kalpana P, ``An energy-efficient XOR gate implementation resistant to power analysis attacks'', {\it J. Engr. Sci. Technol.}, vol. 10, 1275-1292 (2015).
\bibitem{ryan}
R. Keech, L. Ye, J. L. Bosse, G. Esteves, J. Guerrier, J. L. Jones, M. L. Kuroda, B. D. Huey and S. Trolier-McKinstry, ``Declamped Piezoelectric Coefficients in Patterned 70/30
Lead Magnesium Niobate–Lead Titanate Thin Films'', {\it Adv. Funct. Mater.}, vol. 27, Art. 1605014 (2017). https://doi.org/10.1002/ADFM.201605014.
\bibitem{suman}
S. Datta, W. Chakrabarty and M. Radosavljevic, ``Toward attojoule switching energy in logic transistors'', {\it Science}, vol. 378, 733-740 (2022). https://doi.org/10.1126/science.ade7656.
\bibitem{wiki}
$https://en.wikipedia.org/wiki/XOR-gate$. Accessed March 3, 2026.

\end{thebibliography}
\end{document}